\documentclass[12pt]{article}
\pdfoutput=1

\usepackage{amsmath} 
\usepackage{amssymb} 
\usepackage{graphicx}
\usepackage{url}
\usepackage{color}
\usepackage{hyperref}
\definecolor{refcolor}{RGB}{0,0,190}
\hypersetup{
    colorlinks,
    citecolor=refcolor,
    filecolor=refcolor,
    linkcolor=refcolor,
    urlcolor=refcolor
}

\def\({\left(}
\def\){\right)}

\newcommand{\op}[1]{\operatorname{#1}}

\newcommand{\de}{\op{d}}

\newcommand{\tr}{\textnormal{tr }}

\newcommand{\ds}{\displaystyle}

\newcommand{\ie}{\textit{i.e.}}

\newcommand{\eg}{\textit{e.g.}}

\newcommand{\sref}[1]{\S\ref{#1}}

\newcommand{\dsfrac}[2]{\ds{\frac{#1}{#2}}}

\newcommand{\schw}{Schwarzschild}
\newcommand{\rn}{Reissner-Nordstr\"om}
\newcommand{\kn}{Kerr-Newman}

\newcommand{\image}[3]{
\begin{figure}[ht]
\begin{center}
\includegraphics[width=#2\textwidth]{#1}
\caption{\small{\label{#1}#3}}
\end{center}
\end{figure}
}

\begin{document}

\author{Ovidiu Cristinel Stoica
\thanks{Department of Theoretical Physics, National Institute of Physics and Nuclear Engineering -- Horia Hulubei, Bucharest, Romania. Email: \href{mailto:cristi.stoica@theory.nipne.ro}{cristi.stoica@theory.nipne.ro},  \href{mailto:holotronix@gmail.com}{holotronix@gmail.com}}
}

\title{Revisiting the black hole entropy and the information paradox}
\maketitle

\begin{abstract}
The black hole information paradox and the black hole entropy are currently extensively researched. The consensus about the solution of the information paradox is not yet reached, and it is not yet clear what can we learn about quantum gravity from these and the related research. It seems that the apparently irreducible  paradoxes force us to give up on at least one well-established principle or another. Since we are talking about a choice between the principle of equivalence from general relativity, and some essential principles from quantum theory, both being the most reliable theories we have, it is recommended to proceed with caution and search more conservative solutions. These paradoxes are revisited here, as well as the black hole complementarity and the firewall proposals, with an emphasis on the less obvious assumptions. Some arguments from the literature are reviewed, and new counterarguments are presented. Some less considered less radical possibilities are discussed, and a conservative solution, which is more consistent with both the principle of equivalence from general relativity and the unitarity from quantum theory, is discussed.
\end{abstract}

\section{Introduction}

By applying general relativity and quantum field theory on curved spacetime, Hawking arrives at the conclusion that the information is lost in the black holes, and this breaks the predictability \cite{Haw76}. Apparently, no matter how was formed and what information was contained in the matter falling in a black hole, the only degrees of freedom characterizing it are its mass, angular momentum, and electric charge, so black holes are ``hairless'' \cite{Israel1967NoHairSchwarzschild,Israel1968NoHairElectrovac,Carter1971NoHairKerr,MTW1973Gravitation}. This means that the information describing the matter crossing the event horizon is lost, because nothing outside the black hole reminds us of it. In general relativity this information loss is irreversible, not only because we cannot extract it from beyond the event horizon, but also because in a finite time the infalling matter reaches the singularity of the black hole. And the occurrence of singularities is unavoidable, according to the singularity theorems \cite{Pen65,HP70,HE95}. This already seemed to be a problem, but it would not be so severe if we at least know that the information is still there, censored behind the horizon \cite{Pen69,Pen98}. But we are not even left with this possibility, since Hawking proved that quantum effects make the black holes evaporate \cite{Haw75ParticleCreation}. It was already expected that black holes should radiate, after the realization that they have entropy and temperature \cite{Bekenstein1973BHEntropy,Haw75ParticleCreation}, and these should be part of an extension of thermodynamics which includes matter as well. This evaporation is thermal, and after the black hole reaches a planckian size, it explodes and reveals to the exterior world that the information is indeed lost. In addition, if the quantum state prior to the formation of the black hole was pure, the final state is mixed, increasing the drama even more. Moreover, a problem seems to occur long before the complete evaporation, since the black hole entropy seems to increase during evaporation, until the Page time is reached \cite{Page1993AverageEntropy}. Some consider this to be the real black hole information paradox \cite{Marolf2017BlackHoleInformationProblem}.

Mainly for general relativists the information loss seemed to be definitive and yet not a big problem \cite{UnruhWald2017InformationLoss}, position initially endorsed by Hawking too. On the other hand, for high energy physicists, loss of unitarity was considered a problem, and various proposals to fix it appeared (see for example \cite{preskill1993bh-info,Giddings1995BlackHoleInformationParadox,HossenfelderSmolin2010ConservativeBlackHoleInformation} and references therein). For example, \emph{remnants} were proposed, containing the information remaining in the black hole after evaporation. The remnant is in a mixed state, but together with the Hawking radiation form a pure state. A possible cause for remnants are the yet unknown quantum corrections expected to occur when the black hole becomes too small, comparable to the Plank scale, and the usual analysis of Hawking radiation no longer applies \cite{Hawking1982UnpredictabilityOfQuantumGravity,Giddings1994ConstraintsBlackHoleRemnants,Giddings1995BlackHolesNotInfinitelyProduced}.
There are other possibilities, some being discussed in the above mentioned reviews. For example it was proposed that the information leaks out of the black hole through evaporation -- including by quantum tunneling, that it escapes at the final explosion, or that it leaks out of the universe in a baby universe \cite{Markov1984BabyUniverse} \cite{Parikh2000HawkingRadiationTunneling}. Another possibility is that the information escapes as Hawking radiation by \emph{quantum teleportation} \cite{Lloyd2006EscapeBlackHole}, which actually happens as if the particle zig-zags forward and backward in time to escape without exceeding the speed of light. This is not so unnatural, if we assume that the final boundary condition at the future singularity of the black hole forces the maximally entangled particles to be in a singlet state. There are also \emph{bounce scenarios} \cite{Frolov2014BHInfo}, or by using local scale invariance to avoid singularities \cite{Dominis2016BHInfoLocalScaleInvariance}. Some bounce scenarios are based on \emph{loop quantum gravity}, like \cite{Ashtekar2008BlackHoleInformationBounce2D,Ashtekar2011BlackHoleEvaporation2D}, as well as \emph{black hole to white hole tunneling scenarios} in which quantum tunneling is supposed to break the Einstein equation, and the apparent horizon is prevented to evolve into an event horizon \cite{Rovelli2014PlanckStars,HaggardRovelli2015BlackHoleFireworks}. It would take a long review to do justice to the various proposals, and this is beyond the scope of this article.

The dominating proposed solution was, for two decades, \emph{black hole complementarity} \cite{Susskind1993StretchedHorizonBlackHoleComplementarity,StephenstHooft1994black,Susskind2004BlackHoleInformation}. This was later challenged by the \emph{firewall paradox} \cite{AMPS2013BlackHolesFirewalls}. The debate is not settled down yet, but the dominant opinion seems to be that we have to give up of at least one principle considered fundamental so far, and the unlucky one is most likely the principle of equivalence from general relativity. One of the objectives of the present article is to show that we can avoid this radical solution while keeping unitarity.

The problems related to the black hole information loss are considered important, being seen as a benchmark for the candidate theories of quantum gravity, which are expected to solve these problems.

The main purpose of this discussion is to identify the main assumptions, and see if it is possible to solve the problem in a less radical way. I argue that some of the usually made assumptions are unnecessary, that there are less radical possibilities, and that the black hole information problem is not a decisive test for candidate theories of quantum gravity. New counterarguments to some popular models proposed in relation to the black hole information problem are the following. 
Black hole complementarity is discussed in Section \sref{s:bh-complementarity}, in particular that an argument by Susskind, aiming to prove that no-cloning is satisfied by the black hole complementarity, does not apply to most black holes \sref{s:bh-complementarity-cloning}, that its main argument, the ``no-omniscience'' proposal, does not really hold for black holes in general \sref{s:bh-complementarity-omniscience}, and that black hole complementarity is also at odds with the principle of equivalence \sref{s:bh-complementarity-cloning-pp-equivalence}.
As for the firewall proposal, in Section \sref{s:bh-firewall-unitarity} I explain why the tacit assumption that unitarity should apply to the exterior of the black hole and we should ignore the interior is not justified, and anyway if taken as true, it imposes boundary conditions to the field, which is why the firewall seems to emerge.
Section \sref{s:bh-entropy} is dedicated to black hole entropy. In Section \sref{s:bh-entropy-time-reversal} I present an argument based on time symmetry that the true entropy is not necessarily proportional to the area of the event horizon, at best in the usual cases is bounded. This has negative implications to the various proposals that the event horizon would contain some bits representing the microstates of the black hole, discussed in Section \sref{s:bh-entropy-counting}. This may also explain the so-called ``real black hole information paradox'', discussed in Section \sref{s:bh-real-paradox}. Section \sref{s:bh-entropy-bh-laws} contains an explanation of the fact that if the laws of black hole mechanics should be connected with those of thermodynamics, this happens already at the classical level, so they are not necessarily indications of quantum gravity or tests of such approaches. Section \sref{s:bh-hair} contains arguments that one should not read too much in the so-called no-hair theorems, in particular they do not constrain, contrary to a widespread belief, neither the horizon nor the interior of a black hole. A major motivation invoked for the theoretical research of the black hole information and entropy is that these may provide a benchmark to test approaches to quantum gravity, but in Section \sref{s:bh-entropy-benchmark} I argue that these features appear merely by considering quantum fields on spacetime. Consequently, any approach to quantum gravity which includes both quantum field theory and the curved spacetime of general relativity, as a minimal requirement, will also satisfy the consequences derived from them.

To my knowledge, the above mentioned arguments, presented in more detail in the following, are new, and in the cases when I was aware of other results seeming to point in the same direction, I gave the relevant references.
While most part of the article may look like a review of the literature, it is a critical review, aiming to point out some assumptions which, in my opinion, drove us too far from the starting point, which is just the most straightforward and conservative combination of quantum field theory with the curved background of general relativity. The entire structure of arguments converges therefore towards a more conservative picture than that suggested by the more popular proposals. The counterarguments are meant to build up the willingness to consider the less radical proposal that I made, which follows naturally from my work on singularities in standard general relativity (\cite{Sto13a} and references therein), and is discussed in Section \sref{s:bh-conservative}. The background theory is presented in \sref{s:sgr}, and a new, enhanced version of the proposal, is made in \sref{s:sgr-time-evol}.

\section{Black hole evaporation}

Hawking's derivation of the black hole evaporation \cite{Haw75ParticleCreation,Haw76} has been disputed and checked many times, redone in different settings, and it turned valid, at most allowing some improvements of the unavoidable approximations, and mild generalizations. But the result is correct, the radiation is as predicted, and thermal in the Kubo-Martin-Schwinger sense \cite{Kubo1957Thermal,Martin1959Thermal}. Moreover, it is corroborated via the principle of equivalence with the Unruh radiation, which takes place in the Minkowski spacetime for accelerated observers \cite{Unruh1976UnruhEffect}. Hawking's derivation is obtained in the framework of quantum field theory on curved spacetime, but since the black hole is considered large, and the time scale is also large, the spacetime curvature induced by the radiation is ignored.

The derivation, as well as the discussion surrounding black hole information, require the framework of quantum field theory on curved spacetime \cite{Fulling1973nonuniqueQFTCurved,Davies1975scalarProductionQFTCurved,Wal94}. Quantum field theory on curved spacetime is a good effective limit of the true but yet unknown theory of quantum gravity. On curved background there is no Poincar\'e symmetry to select a preferred vacuum, so there is no canonical Fock space construction of the Hilbert space. 
The stress-energy expectation value of the quantum fields, $\langle\hat T_{ab}(x)\rangle$, is connected with the spacetime geometry via Einstein's equation,
\begin{equation}
\label{eq:einstein-qft}
R_{ab}-\frac12 Rg_{ab}+\Lambda g_{ab} = \frac{8\pi G}{c^4}\langle\hat T_{ab}(x)\rangle,
\end{equation}
where $R_{ab}$ is the \emph{Ricci tensor}, $R$ is the \emph{scalar curvature}, $g_{ab}$ is the \emph{metric tensor}, $\Lambda$ is the \emph{cosmological constant}, $G$ is Newton's gravitational constant, and $c$ is the speed of light constant.

But in the calculations of the Hawking radiation the gravitational backreaction is ignored, being very small. To have well behaved solutions, the spacetime slicing is such that the intrinsic and extrinsic curvatures of the spacelike slices are considered small compared to the Plank length; the curvature in a neighborhood of the spacelike surface is also taken to be small. The wavelengths of particles are considered large compared to the Plank length. The energy and momentum densities are assumed small compared to the Plank density. The stress-energy tensor satisfies the positive energy conditions. The solution evolves smoothly into future slices that also satisfy these conditions. 

The canonical (anti-)commutation relations at distinct points of the slice are imposed. It is assumed a decomposition into positive and negative frequency solutions, to which the Fock construction is applied to obtain the Hilbert space. The renormalizability of the stress-energy expectation value $\langle\hat T_{ab}(x)\rangle$, and the uniqueness of the $n$-point function $\langle\hat\phi(x_1)\ldots\hat\phi(x_n)\rangle$, are ensured by imposing the \emph{Hadamard condition} to the quantum states \cite{Wal94}. This condition is needed because when two of the $n$-points coincide, there is no invariant way to define the $n$-point function on curved spacetime. The Hadamard condition is imposed on the \emph{Wightman function} $G(x,y)=\langle\hat\phi(x)\hat\phi(y)\rangle$, and it is preserved under time evolution. This condition is naturally satisfied in the usual quantum field theory in Minkowski spacetime. It ensures the possibility to renormalize the stress-energy tensor, and to prevent it to diverge.

The Fock space construction of the Hilbert space can be made in many different ways in curved spacetime, since the decomposition into positive and negative frequency solutions depends on the choice of the slicing of spacetime into spacelike hypersurfaces.

Suppose that a basis of annihilation operators is $(\hat a_\nu)$, and they satisfy the canonical commutation relations if they are bosons, and the canonical anticommutation relations if they are fermions. Another observer has a different basis of annihilation operators $(\hat b_\omega$), assuming that the spacetime is curved, or that one observer accelerates with respect to the other. The two bases are related by the \emph{Bogoliubov transformations},
\begin{equation}
\label{eq:bogoliubov}
\hat b_\omega = \frac 1{2\pi} \int_0^\infty \(\alpha_{\omega\nu}\hat a_\nu + \beta_{\omega\nu}\hat a_\nu^\dagger\)\de\nu,
\end{equation}
where $\alpha_{\omega\nu}$ and $\beta_{\omega\nu}$ are the \emph{Bogoliubov coefficients}.

The Bogoliubov transformation preserves the canonical (anti)commutation relations and expresses the change of basis of the Fock space, allowing us to move from one construction to another. The Bogoliubov transformations are linear but not unitary. They are symplectic for bosons and orthogonal for fermions though. The number of particles is not preserved, so there is no invariant notion of particles. 

This is in fact the reason for both the Unruh effect near a Rindler horizon, and the Hawking evaporation near a black hole event horizon.
Because of the nonunitarity of the Bogoliubov transformation relating the Fock space representations of two distinct observers, particles can be produced \cite{Fulling1973nonuniqueQFTCurved,Davies1975scalarProductionQFTCurved,Unruh1976UnruhEffect}, including for black holes \cite{Haw75ParticleCreation}. This means that what is a vacuum state for an inertial observer, for an accelerated one is a state with many particles. This is true in the Minkowski spacetime, if one observer is accelerated with respect to the other, but also for two inertial observers, if the curvature is relevant, as in the case of infalling and escaping observers near a black hole. Moreover, the many-particle state in which the vacuum of one observer appears to the other is thermal.
The particle and the antiparticle created in pair during the evaporation are maximally entangled.


\section{Black hole complementarity}
\label{s:bh-complementarity}

While Hawking's derivation of the black hole evaporation is rigorous and the result is correct, the implication that the information is definitively lost can be challenged.
In fact, most of the literature on this problem is trying to find a workaround to restore the lost information and the unitarity. The most popular proposals like black hole complementarity and firewalls do not actually dispute the calculations, but rather they add the requirement that the Hawking radiation should contain the complete information.

Additional motivation for unitarity comes from the AdS/CFT correspondence \cite{Maldacena1999Duality}. The AdS/CFT is not yet rigorously proven, and it is in fact against the current cosmological observations that the cosmological constant is positive \cite{RIE98,PER99}, but it is widely considered true or standing for a correct gauge-gravity duality, and it is likely that it convinced Hawking to change his mind about information loss \cite{Hawking2005InformationLoss}.

The favorite scenario among high energy physicists was, for two decades, the idea of \emph{black hole complementarity} \cite{Susskind1993StretchedHorizonBlackHoleComplementarity,StephenstHooft1994black,Susskind2004BlackHoleInformation}, which supposedly resolves the conflict between unitarity, essential for quantum theory, and the principle of equivalence from general relativity.
Susskind and collaborators framed the black hole information paradox as implying a contradiction between unitarity and the principle of equivalence.
They proposed a radical solution of this apparent conflict, by admitting two distinct Hilbert space descriptions for the infalling matter and the escaping radiation \cite{Susskind1993StretchedHorizonBlackHoleComplementarity}. 

Assuming that unitarity is to be restored by evaporation alone, the infalling information should be found in the Hawking radiation, or should somehow remain above the black hole event horizon -- forming the \emph{stretched horizon} \cite{Susskind1993StretchedHorizonBlackHoleComplementarity}, similar to the \emph{membrane paradigm} \cite{PriceThorne1986MembraneParadigm}). But since this information falls in the black hole, it would violate the \emph{no cloning theorem} of quantum mechanics \cite{Park1970NoCloning,WoottersZurek1982NoCloning,Dieks1982NoCloning}. If the cloning does not happen, either the information is not recovered (and unitarity is violated) or no information can cross the horizon, which would violate the principle of equivalence from general relativity, which implies that nothing dramatic should happen at the event horizon, assuming that the black hole is large enough. The black hole complementarity assumes that both unitarity and the principle of equivalence hold true, by allowing cloning, but the cloning cannot be observed, because each observer sees only one copy. The infalling copy of the information is accessible to an infalling observer only (usually named Alice), and the escaping one to an escaping observer (Bob). Susskind and collaborators conjectured that Alice and Bob can never meet to confirm that the infalling quantum information was cloned and the copy escaped the black hole.

At first sight it may seem that the black hole complementarity solves the contradiction by allowing it to exist, as long as no experiment is able to prove it. Alice and Bob's lightcones intersect, but none of them is included in the other, and they cannot be made so. This means that whatever slicing of spacetime they choose in their reference frames, the Hilbert space constructions they make will be different. So it would be impossible to compare quantum information from the interior of the black hole with the copy of quantum information escaping it. And it is impossible to conceive an observer able to see both copies of information -- this would be the so-called \emph{omniscience condition}, which is rejected by Susskind and collaborator to save both unitarity and the principle of equivalence.

\subsection{No-cloning and timelike singularities}
\label{s:bh-complementarity-cloning}

An early objection to the proposal that Alice and Bob can never compare the two copies of quantum information was that the escaping observer Bob can collect the escaping copy of the information, and jump into the black hole to collect the infalling copy. This objection was rejected because in order to collect a single bit of infalling information from the Hawking radiation, Bob should wait until the black hole loses half of its initial mass by evaporation -- the time needed for this to happen is called the \emph{Page time} \cite{Page1993AverageEntropy}. So if Bob decides to jump in the black hole to compare the escaping information with the infalling one, it would be too late, because the infalling information will have just enough time to reach the singularity.

The argument based on the Page time works well, but it applies only to black holes of the {\schw} type (more precisely this is an Oppenheimer-Snyder black hole \cite{OS1939collapse}), whose singularity is a spacelike hypersurface. For rotating or electrically charged black holes, the singularity is a timelike curve or cylinder. In this case, Alice can carry the infalling information around the singularity for an indefinitely long time, without reaching the singularity. So Bob will be able to reach Alice and confirm that the quantum information was cloned.

This objection is relevant, because for the black hole to be of {\schw} type, two of the three parameters defining the black hole, the angular momentum and the electric charge, have to vanish, which is very unlikely. The things are even more complicated if we take into account the fact that during evaporation or any additional particle falling in the black hole, the type of the black hole changes. Usually particles have non-vanishing electric charges and spin, and even if an infalling particle is electrically neutral and has the spin equal to $0$, most likely it will not collide with the black hole radially. This continuous change of the type of the black hole may result in changes of type of the singularity, rendering the argument based on the Page time invalid.

In subsection \sref{s:bh-complementarity-omniscience} we will see that even if the black hole somehow manages to remain of {\schw} type, the cloning can be made manifest to a single observer.

\subsection{No-cloning and the principle of equivalence}
\label{s:bh-complementarity-cloning-pp-equivalence}

Because of the principle of equivalence, Susskind's argument should also hold for Rindler horizons in Minkowski spacetime. The equivalence implies that Bob is an accelerated observer, and Alice is an inertial observer, who crosses Bob's Rindler horizon. Because of the Unruh effect, Bob will perceive the vacuum state as thermal radiation, while for Alice it would be just vacuum. Bob can see Alice being burned at the Rindler horizon by the thermal radiation, but Alice will experience nothing of this sort. But since they are now in the Minkowski spacetime, Bob can stop and go back to check the situation with Alice, and he will find that she did not experience the thermal bath he saw her experiencing. While we can just say that the complementarity should be applied only to black holes, to rule it out for the Rindler horizon and still maintain the idea of stretched horizon only for black holes, this would be at odds with the principle of equivalence which black hole complementarity is supposed to rescue.

\subsection{The ``no-omniscience'' proposal}
\label{s:bh-complementarity-omniscience}

The resolution proposed by black-hole complementarity appeals to the fact that the Hilbert spaces constructed by Alice and Bob are distinct, which would allow quantum cloning, as long as the two copies belong to distinct Hilbert spaces and there is no observer to see the violation of the no-cloning theorem.
This means that the patches of spacetime covered by Alice and Bob are distinct, such that apparently no observer can cover both of them. If there were such an ``omniscient'' observer, he or she would see the cloning of quantum information, and see that the laws of quantum theory are violated.

Yet, there is such an observer, albeit is moving backwards in time, see fig. \ref{bh-info-complementarity-omniscience.png}. Remember that the whole point of trying to restore the loss information and unitarity is because quantum theory should be unitary. This means not only deterministic, but also that the time evolution laws have to be time symmetric, as quantum theory normally is, so that we can recover the lost information. So everything quantum evolution does forward in time, it should be accessible by backwards in time evolution. An observer going backwards in time, Charlie, can then in principle be able to perceive both copies of the information carried by Alice and Bob, so he is ``omniscient''.

\image{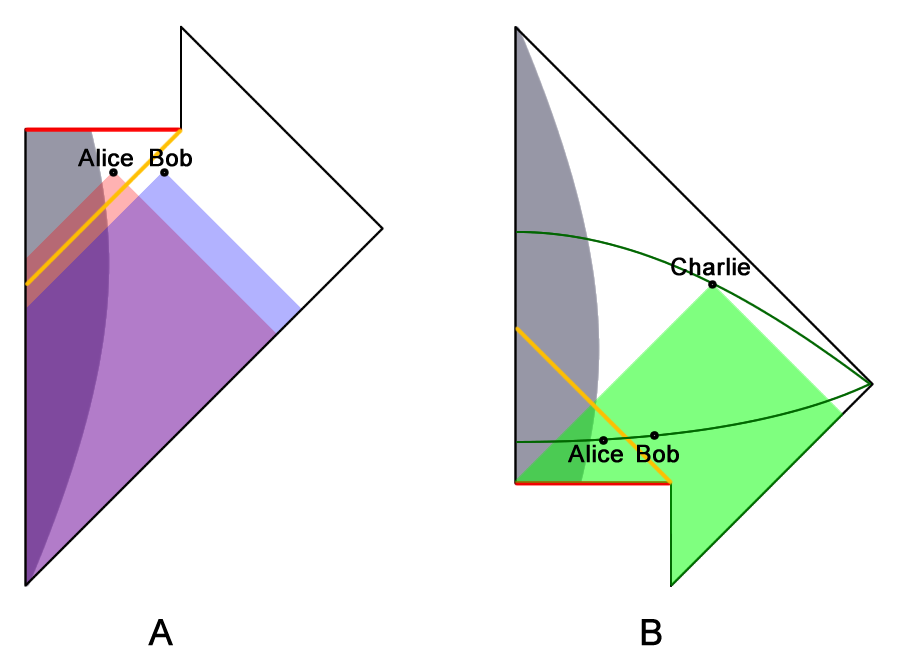}{0.9}{\textbf{A.} The Penrose diagram of black hole evaporation, depicting Alice and Bob and their past lightcones. \textbf{B.} The Penrose diagram of a backwards in time observer Charlie, depicting how he observes Alice and Bob, and the quantum information each of them caries, even if this information is cloned, therefore disclosing a violation of quantum theory.}

One can try to rule Charlie out, on the grounds that he violates causality, or more precisely the second law of thermodynamics \cite{schulman1997timeArrowsAndQuantumMeasurement}. But from the point of view of quantum theory, the von Neumann entropy is preserved by unitary evolution, and the quantum evolution is reversible anyway, so it is irrelevant if in our real universe there is a thermodynamic arrow of time, this does not invalidate a principial thought experiment like this one.

\section{The firewall paradox}
\label{s:bh-firewall}

After two decades since the proposal of black hole complementarity, this solution was disputed by the \emph{firewall paradox} \cite{AMPS2013BlackHolesFirewalls}, which suggested that the equivalence principle should be violated at the event horizon, where a highly energetic curtain or a singularity should form to prevent the information falling inside the black hole.

The firewall argument takes place in the same settings as the black hole complementarity proposal, but this time it involves the \emph{monogamy of entanglement}. More precisely, it is shown that the late radiation has to be maximally entangled with both the early radiation and with the infalling counterpart of the late radiation. Since the monogamy of entanglement forbids this, it is proposed that one of the assumptions has to go, most likely the principle of equivalence. The immediate reaction varied from quick acceptance, to arguments that the paradox is solved too by the black hole complementarity \cite{Bousso2012Complementarityv1,HarlowHayden2013quantum}. After all, we can think of the late radiation as being entangled with the early one in Bob's Hilbert space, and with the infalling radiation in Alice's Hilbert space. But it turned out that, unlike the case of the violation of the no-cloning theorem, the violation of monogamy cannot be resolved by Alice and Bob having different Hilbert spaces \cite{Bousso2012Complementarityv2}.

One can argue that if the firewall experiment is performed, it creates the firewall, and if it is not performed, Alice sees no firewall, so black hole complementarity is not completely lost. Susskind and Maldacena proposed the \emph{ER=EPR} solution, which states that if entangled particles are thrown in different black holes, then they become connected by a wormhole \cite{MaldacenaSusskind2013EREPR}, also see \cite{BryanMedved2017BHInfo}. The firewall idea also stimulated various discussion about the relevance of complexity of quantum computation, and error correction codes, in the black hole evaporation and decoding the information from the Hawking radiation using unitary operations (see \cite{HarlowHayden2013quantum}, \cite{StanfordSusskind2014Complexity}, and \cite{Aaronson2016ComplexityBlackHoles} and references therein).

Various proposals to rescue both the principle of equivalence and unitarity were made, for example based on the entropy of entanglement across the event horizon in \cite{Braunstein2013InfoBH}, \cite{YosifovFilipov2017EntropicBHInfo}. Hawking proposed that the black hole horizons are only apparent horizons and never actual event horizons \cite{Hawking2014BHInfoWeatherForecasting}. Later, Hawking proposed that supertranslations allow the preservation of information, and further expanded the idea with Perry and Strominger \cite{Hawking2015SupertranslationsBHInfo,Hawking2016SofthairBHInfo,Hawking2017SuperrotationSupertranslationBHInfo}.

Having to give up the principle of equivalence or unitarity is a serious dilemma, so it worth revisiting the arguments to find a way to save both.

\subsection{The meaning of ``unitarity''}
\label{s:bh-firewall-unitarity}

In the literature about black hole complementarity and firewalls, by the assumption or requirement of ``unitarity'', we should understand ``unitarity of the Hawking radiation'', or more precisely, ``unitarity of the quantum state exterior to the black hole''. Let us call this \emph{exterior unitarity}, to emphasize that it ignores the interior of the black hole. It is essential to clarify this, because when we feel that we are forced to choose between unitarity and the principle of equivalence, we are in fact forced to choose between exterior unitarity and the principle of equivalence. This assumption about is also at the origin of the firewall proposal. So no choice between unitarity and the principle of equivalence is enforced to us, unless by ``unitarity'' we understand ``exterior unitarity''. 

The idea that unitarity should be restored from the Hawking radiation alone, ignoring the interior of the black hole, was reinforced by the holographic principle and the idea of stretched horizon \cite{Susskind1993StretchedHorizonBlackHoleComplementarity,Susskind1995HolographicPrinciple,StephenstHooft1994black} -- a place just above the event horizon which presumably stores the infalling information until it is restored through evaporation -- and it was later reinforced even more by the AdS/CFT conjecture \cite{Maldacena1999Duality}. But it is not excluded to solve the problem by taking into consideration both the exterior and interior of the black hole and the corresponding quantum states. 
A proposal accounting for the interior in the AdS/CFT correspondence, based on the impossibility to localize the quantum operators in quantum gravity in a background-independent manner, was made in \cite{Papadodimas2014BHInteriorHolographic}. A variation of the AdS/CFT leading to a regularization was made in \cite{Wang2017BHInfo}

In fact, considering both the exterior and the interior of the black hole is behind proposals like remnants and baby universes. But we will see later that there is a less radical option.

Exterior unitarity, or the proposal that the full information and purity are restored from Hawking radiation alone, simply removes the interior of the black hole from the reference frame of an escaping observer, consequently from his Hilbert space. This type of unitarity imposes a boundary condition to the quantum fields, which is simply that there is no relevant information inside the black hole. So it is natural that at the boundary of the support of the quantum fields, which is the black hole event horizon, quantum fields behave as if there is a firewall. This is what the various estimates revealing the existence of a highly energetic firewall or horizon singularity confirm. 
Note that since the boundary condition which aims to rescue the purity of the Hawking radiation is a condition about the final state, sometimes its consequences give the impression of a conspiracy, as sometimes Bousso and Hayden put it \cite{Gefter2014ComplexityOnTheHorizon}.

While I have no reason to doubt the validity of the firewall argument  \cite{AMPS2013BlackHolesFirewalls}, I have reservations about assuming unitarity as referring only to quantum fields living only to the exterior of the black hole, while ignoring those from its interior.

\subsection{Firewalls versus complementarity}
\label{s:bh-firewall-complementarity}

The initial Hilbert spaces of Alice and Bob are not necessarily distinct. Even if they are and their Fock constructions are distinct, each state from one of the spaces may correspond to a state from the other. The reason is that a basis of annihilation operators in Alice's frame, say $(\hat a_\nu)$, is related to a basis of annihilation operators in Bob's frame, $(\hat b_\omega)$, by a Bogoliubov transformation \eqref{eq:bogoliubov}. The Bogoliubov transformation is linear, although not unitary.

Thus, one may hope that the Hilbert spaces of Alice and Bob may be identified, even though through a very scrambled vector space isomorphism, so that black hole complementarity saves the day.
However, exterior unitarity imposes that the evolved quantum fields from the Hilbert spaces have different supporting regions in spacetime. While before the creation of the black hole they may have the same support in the spacelike slice, they evolve differently, because of the exterior unitarity condition. Bob's system evolves so that his quantum fields are constrained to the exterior of the black hole, while Alice's quantum fields include the interior too. Bob's Hilbert space is different, because when the condition of exterior unitarity was imposed, it excluded the interior of the black hole. 
So even if the initial underlying vector space is the same for both the Hilbert space constructed by Alice and that constructed by Bob, their coordinate systems diverged in time, so the way they slice spacetime became different. While normally Alice's vacuum is perceived by Bob as loaded with particles in a thermal state, this time in Bob's frame Alice's vacuum energy becomes singular at the horizon.
This makes the firewall paradox a problem for black hole complementarity. A cleaner argument based on purity rather than monogamy is made by Bousso \cite{Bousso2013FirewallsDoublePurity}.

An interesting issue is that Bob can infer that if the modes he detects passed very close to the event horizon, they were redshifted. So evolving the modes backwards in time, it must be that the particle passes close to the horizon at a very high frequency, maybe even higher than the Plank frequency. Does this mean that Alice should feel dramatically this radiation? There is the possibility that for Alice, Bob's high frequency modes are hidden in her vacuum state. This is also confirmed by acoustic black holes \cite{Weinfurtner2011AcousticBlackHoles}. Only if these modes are somehow disclosed, for example if Bob, being accelerated, performs some temperature detection nearby Alice, these modes may become manifest due to the projection postulate, otherwise they remain implicit in Alice's vacuum.

It seems that the strength of the firewall proposal comes from rendering black hole complementarity unable to solve the firewall paradox. They are two competing proposals, both aiming to solve the same problem. While one can logically think that proposals which take into account the interior of black holes to restore unitarity are good candidates as well, and that they may have the advantage of rescuing the principle of equivalence, sometimes they are dismissed as not addressing the ``real'' black hole information paradox. I will say more about this in Section \sref{s:bh-real-paradox}.

\section{Black hole entropy}
\label{s:bh-entropy}

The purposes of this Section are to prepare for Section \sref{s:bh-real-paradox}, and to discuss the implications of black hole entropy for the black hole information paradox, and for quantum gravity.

The entropy bound of a black hole is proportional to the area of the event horizon \cite{Bekenstein1973BHEntropy,BCH1973fourLawsBH,Bousso2002HolographicPrinciple},
\begin{equation}
\label{eq:bh-entropy}
S_{BH} = \frac{k_B A}{4\ell_P^2},
\end{equation}
where $k_B$ the Boltzmann constant, $A$ the area of the event horizon, and $\ell_P$ the Plank length.

The black hole entropy bound \eqref{eq:bh-entropy} was suggested by Hawking's result that the black hole horizon area never decreases \cite{Haw71BlackHoleArea}, as well as the development of this result into the four laws of black hole mechanics \cite{BCH1973fourLawsBH}.

\subsection{The area of the event horizon and the entropy}
\label{s:bh-entropy-time-reversal}
 
It is tempting to think that the true entropy of quantum fields in spacetime should also include the areas of the event horizons. In fact, there are computational indications that the black hole evaporation leaks the right entropy to compensate the decrease of the area of the black hole event horizon.

But there is a big difference between the entropy of quantum fields, and the areas of horizons. First, entropy is associated to the state of the matter (including radiation, of course). If we look at the phase space, we see that the entropy is a property of the state alone, so it is irrelevant if the system evolves in one direction of time or the opposite, the entropy corresponding to the state at a time $t$ is the same. The same is true for quantum entropy, associated to the quantum states, which in fact is preserved by unitary evolution and is the same in either time direction.

On the other hand, the very notion of event horizon in general relativity depends on the direction of time. By looking again at fig. \ref{bh-info-complementarity-omniscience.png} \textbf{B}, this time without being interested in black hole complementarity, we can see that for Charlie there is no event horizon. But the entropy corresponding to matter is the same independently of his time direction. So even if we are able to put the area on the event horizon in the same formula with the entropy of the fields and still have the second law of thermodynamics, the two terms behave completely differently. So if the area of the event horizon is required to compensate for the disappearance of entropy beyond the horizon, and for its reemergence as Hawking radiation, for Charlie the things are quite different, because he has full clearance to the interior of the black hole, which for him is white. In other words, he is so omniscient that he knows the true entropy of the matter inside the black hole, and not a mere bound given by the event horizon.

This is consistent with the usual understanding of entropy as hidden information -- indeed, the true information about the microstates is not accessible, only the macrostate, and this is what entropy stands for. But it is striking, nevertheless, to see that black holes do the same, yet in a completely time-asymmetric manner. This is because the horizon entropy is just a bound for the entropy beyond the horizon, the true entropy is a property of the state.

\subsection{Black hole mechanics and thermodynamics -- matter or geometry?}
\label{s:bh-entropy-bh-laws}

The four laws of black hole mechanics are \cite{BCH1973fourLawsBH,Mann2015BlackHolesThermodynamicsInformationFirewalls}:
\begin{itemize}
	\item 
\textbf{0th Law.} The surface gravity $\kappa$ is constant over the event horizon.
	\item 
\textbf{1st Law.} For nearby solutions, the differences in are equal to differences in area times the surface gravity, plus some additional terms similar to work.
	\item 
\textbf{2st Law.} In any physical process, the area of the event horizon never decreases (assuming positive energy of matter and regularity of spacetime).
	\item 
\textbf{3rd Law.} There is no procedure, consisting of a finite number of steps, to reduce the surface gravity to zero.
\end{itemize}

The analogy between the laws of black hole mechanics and thermodynamics is quite impressive \cite{Mann2015BlackHolesThermodynamicsInformationFirewalls}. In particular, enthalpy, temperature, entropy, and pressure, correspond respectively to the mass of the black hole, its surface gravity, its horizon area, and the cosmological constant.

These laws of black hole mechanics are obtained in purely classical general relativity, but were interpreted as laws of black hole thermodynamics \cite{Haw75ParticleCreation}, \cite{parker1969ParticleCreation}, \cite{dolan2012BlackHoleThermodynamicsPressure}. Their thermodynamical interpretation occurs when considering quantum field theory on curved spacetime, and it is expected to follow more precisely from the yet to be found quantum gravity.

Interestingly, despite their analogy with the laws of thermodynamics, the laws of black hole mechanics hold in purely classical general relativity. While we expect general relativity to be at least a limit theory of a more complete, quantized one, it is a standalone and perfectly selfconsistent theory. This suggest that it is possible that the laws of black hole mechanics already have thermodynamic interpretation in the geometry of spacetime.
And this turns out to be true, since black hole entropy can be shown to be the Noether charge of the diffeomorphism symmetry \cite{Wald1993BlackHoleEntropyNoetherChargeDiffeomorphism}. This works exactly for general relativity, and it is different for gravity modified so that the action is of higher order in terms of curvature.
In addition, we already know that Einstein's equation can be understood from an entropic perspective which has a geometric interpretation \cite{jacobson1995thermodynamicsSpacetime,verlinde2011entropicGravity}. 

This is not to say that the interpretations of the laws of black hole mechanics in terms of thermodynamics of quantum fields do not hold, because there are strong indications that it does. My point is rather that there is a thermodynamics of the spacetime geometry, which is tied somehow with the thermodynamics of quantum matter and radiation. This connection is probably made via the Einstein's equation, or whatever equation is the one whose classical limit is Einstein's equation.

\subsection{Do black holes have no hair?}
\label{s:bh-hair}

Classically, black holes are considered to be completely described by their mass, angular momentum, and electric charge. This idea is based on the \emph{no-hair theorems}. These results were obtained for the Einstein-Maxwell equations, assuming that the solutions are asymptotically flat and  stationary. While it is often believed that these results hold universally, they are in fact similar to Birkhoff's theorem \cite{Birkhoff1923SchwarzschildUniqueness}, which states that any spherically symmetric solution of the vacuum field equations must be static and asymptotically flat, hence the exterior solution must be given by the Schwarzschild metric.
Werner Israel establishes that the {\schw} solution is the unique asymptotically flat static non-rotating solution of Einstein’s equation in vacuum, under certain conditions \cite{Israel1967NoHairSchwarzschild}. This was generalized to the Einstein-Maxwell equations (electrovac) \cite{Israel1968NoHairElectrovac,Carter1971NoHairKerr,MTW1973Gravitation}, the result being the characterization of static asymptotically flat solutions only by mass, electric charge, and angular momentum. It is conjectured that this result is general, but counterexamples are known \cite{Heusler1996NoHairCounterexamples,Mavromatos1996NoHairCounterexamples}.

In classical general relativity the black holes radiate gravitational waves, and are expected to converge to a no hair solution very fast. If this is true, it happens asymptotically, and the gravitational waves carry the missing information about the initial shape of the black hole horizon, because classical general relativity is deterministic on regular globally hyperbolic regions of spacetime.

Moreover, it is not known what happens when quantum theory is applied. If the gravitational waves are quantized (resulting in gravitons), it is plausible to consider the possibility that quantum effects prevent such a radiation, like in the case of the electron in the atom. Therefore, it is not clear that the information about the infalling matter is completely lost in the black hole, even in the absence of Hawking evaporation. So we should expect at most that black holes converge asymptotically to the simple static solutions, but if they would reach them in finite time, there would be no time reversibility in GR.

Nevertheless, this alone is unable to provide a solution to the information loss paradox, especially since spacetime curvature does not contain the complete information about matter fields. But we see that we have to be careful when we use the no-hair conjecture as an assumption in other proofs.

\subsection{Counting bits}
\label{s:bh-entropy-counting}

While black hole mechanics suggests that the entropy of a black hole is limited by the Bekenstein bound \eqref{eq:bh-entropy}, it is known that the usual classical entropy of a system can be expressed in terms of its microstates,
\begin{equation}
\label{eq:entropy_micro}
S_Q = -k_B\sum_i p_i \ln p_i,
\end{equation}
where $p_i$ denotes the number of microstates which cannot be distinguished because of the coarse graining, macroscopically appearing as the $i$-th macrostate. A similar formula gives the quantum von Neumann entropy, in terms of the density matrix $\rho$,
\begin{equation}
\label{eq:entropy_vonNeumann}
S = -k_B\tr(\rho\ln\rho).
\end{equation}

Because of the \emph{no-hair theorem} (see Section \sref{s:bh-hair}) it is considered that classical black holes can be completely characterized by the mass, angular momentum, and electric charge, at least from the outside. This is usually understood as suggesting that quantum black holes have to contain somewhere, most likely on their horizons, some additional degrees of freedom corresponding to their microstates, so that equation \eqref{eq:bh-entropy} can be interpreted in terms of eqn. \eqref{eq:entropy_micro}.

It is often suggested that there are some horizon microstates, either floating above the horizon but not falling because of a \emph{brick wall} \cite{ZurekThorne1985StatisticalBlackHoleEntropy,Hooft1985QuantumStructureBlackHole,Mann1992BrickWallsBlackHoles}, or being horizon gravitational states \cite{Carlip1999EntropyBlackHole}.

Other counting proposals are based on counting string excited microstates \cite{Strominger1996MicroscopicStringyBlackHoleEntropy,Horowitz1996CountingBlackHoleEntropy,Dabholkar2005SupersymmetricBlackHoleEntropy}. There are also proposals of counting microstates in LQG, for example by using a Chern-Simons field theory on the horizon, and choosing a particular Immirzi parameter \cite{Ashtekar1998LQGBlackHoleEntropy}.

Another interesting possible origin of entropy comes from \emph{entropy of entanglement} resulting by the reduced density matrix of an external observer \cite{Bombelli1986BlackHoleEntropyAsEntanglementEntropy,Srednicki1993EntropyBlackHole}. This is proportional, but for short distances requires renormalization. 

But, following the arguments in Section \sref{s:bh-entropy-time-reversal}, I think that the most natural explanation of black hole entropy seems to be to consider the internal states of matter and gravity \cite{FrolovNovikov1993DynamicalOriginBlackHoleEntropy}. A model of the internal state of the black hole similar to the atomic model was proposed in \cite{Corda2012BlackHoleThermodynamics,Corda2013BlackHoleQuantumSpectrum,Corda2015BohrBlackHole}.
Models based on Bose-Einstein condensates can be found in \cite{Dvali2014BlackHoleEntropy,Casadio2014SelfSustainedBlackHoles,Casadio2015ThermalBECBlackHoles} and references therein.

Since in Section \sref{s:bh-entropy-time-reversal} it was explained that the horizons just hide matter, hence entropy, and are not in fact the carriers of the entropy, it seems to me more plausible that the structure of the matter inside the black hole is just bounded by the Bekenstein bound, and does not point to an unknown microstructure.

\subsection{A benchmark to test quantum gravity proposals?}
\label{s:bh-entropy-benchmark}

The interest in the black hole information paradox and black hole entropy is not only due to the necessity of restoring unitarity. This research is also motivated by testing various competing candidate theories of quantum gravity. Quantum gravity seems to be far from our experimental possibilities, because it is believed to become relevant at very small scales. On the other hand, black hole information loss and black hole entropy pose interesting problems, and the competing proposals of quantum gravity are racing to solve them. The motivation is that it is considered that black hole entropy and information loss can be explained by one of these quantum gravity approaches.

On the other hand, it is essential to remember how black hole evaporation and black hole entropy were derived. The mathematical proofs are done within the framework of quantum field theory on curved spacetime, which is considered a good effective limit of the true but yet to be discovered theory of quantum gravity. The calculations are made near the horizon, they do not involve extreme conditions like singularities or planckian scales, where quantum gravity is expected to take the lead. The main assumptions are:
\begin{enumerate}
	\item 
quantum field theory on curved spacetime
	\item 
the Einstein equation, with the stress-energy tensor replaced by the stress-energy expectation value $\langle\hat T_{ab}(x)\rangle$ (see eqn. \ref{eq:einstein-qft}).
\end{enumerate}

For example, when we calculate the Bekenstein entropy bound, we do this by throwing matter in a black hole, and see how much the event horizon area increases.

These conditions are expected to hold in the effective limit of any theory of quantum gravity.

But since both the black hole entropy and the Hawking evaporation are obtained from the two conditions mentioned above, this means that any theory in which these conditions are true at least in the low energy limit, is also capable to imply both the black hole entropy and the Hawking evaporation. In other words, if a theory of quantum gravity becomes in some limit the familiar quantum field theory, and also describes Einstein's gravity, it should also reproduce the black hole entropy and the Hawking evaporation.

Nevertheless, some candidate theories to quantum gravity do not actually work in a dynamically curved spacetime, being for example defined on flat or AdS spacetime, yet they still are able to reproduce a microstructure of black hole entropy. This should not be very surprising, given that even in nonrelativistic quantum mechanics, quantum systems bounded in a compact region of space have discrete spectrum. So it may be very well possible that these results are due to the fact that even in nonrelativistic quantum mechanics entropy bounds hold \cite{Bekenstein2005EntropyBound}. In flat spacetime, we can think that the number of states in the spectrum is proportional with the volume. However, when we plug in the masses of the particles in the formula for the {\schw} radius (which incidentally is the same as Michell's formula in Newtonian gravity \cite{Schaffer1979MichellBlackHoles}), we should obtain a relation similar to \eqref{eq:bh-entropy}.

The entropy bound \eqref{eq:bh-entropy} connects the fundamental constants usually considered to be characteristic for general relativity, quantum theory, and thermodynamics. This does not necessarily mean that the entropy of the black hole witnesses about quantum gravity. This should be clear already from the fact that the black hole entropy bound was not derived by assuming quantum gravity, but simply from the assumptions mentioned above. It is natural that, if we plug the information and the masses of the particles in the formula for the {\schw} radius, we obtain a relation between the constants involved in general relativity, quantum theory, and thermodynamics. It is simply a property of the system itself, not a witness of a deeper theory.
But of course, if a candidate theory of quantum gravity fails to pass even this test, this may be a bad sign for it.

\section{The real black hole information paradox}
\label{s:bh-real-paradox}

Sometimes it is said that the true black hole information paradox is the one following from Don Page's article \cite{Page1993AverageEntropy}. For example, Marolf considers that here lies the true paradoxical nature of the black hole information, while he calls the mere information loss and loss of purity ``the straw man information problem'' \cite{Marolf2017BlackHoleInformationProblem}. Apparently the black hole von Neumann entropy should increase with one bit for each emitted photon. At the same time its area decreases by losing energy, so the black hole entropy should also decrease by the usual Bekenstein-Hawking kind of calculation.
So what happens with the entropy of the black hole, does it increase or decrease? This problem occurs much earlier in the evolution of the black hole, when the black hole area is reduced to half of its initial value (the \emph{Page time}), so we don't have to wait for the complete evaporation to notice this problem. As Marolf put it \cite{Marolf2017BlackHoleInformationProblem},
\begin{quote}
This is now a real problem. Evaporation causes the black hole to shrink, and thus to reduce its surface area. So $S_{BH}$ decreases at a steady rate.  On the other hand, the actual von Neumann entropy of the black hole must increase at a steady rate. But the first must be larger than the second. So some contradiction is reached at a finite time.
\end{quote}

I think there are some assumptions hidden in this argument. We compare the von Neumann entropy of the black hole calculated during evaporation with the black hole entropy calculated by Bekenstein and Hawking by throwing particles in the black hole. While the proportionality of the black hole entropy with the area of the event horizon has been confirmed by various calculations for numerous cases, the two types of processes are different, so it is natural that they lead to different states of the black hole, hence to different values for the entropy. This is not a paradox, it is just evidence that the entropy contained in the black hole depends on the way it is created, despite the bound given by the horizon. So it seems more natural not to consider that the entropy of the matter inside the black hole reached the maximum bound at the beginning, but rather that it reaches its maximum at the Page time, due to the entanglement entropy with the Hawking radiation. Alternatively, we may still want to consider the possibility of having more entropy in the black hole than the Bekenstein bound allows. In fact, Rovelli made another argument pointing in the same direction that the Bekenstein-Bound is violated, by counting the number of states that can be distinguished by local observers (as opposed to external observers) using local algebras of observables \cite{Rovelli2017BHEntropyBound}. This argument provided grounds for a proposal of a white hole remnant scenario discussed in \cite{Bianchi2018WhiteHolesRemnants}.

\section{A more conservative solution}
\label{s:bh-conservative}

We have seen in the previous sections that some important approaches to the black hole information paradox and the related topics assume that the interior of the black hole is irrelevant or does not exist, and the event horizon plays the important role. I also presented arguments that if it is to recover unitarity without losing the principle of equivalence, then the interior of the black hole should be considered as well, and the event horizon should not be endowed with special properties. More precisely, given that the original culprit of the information loss is its supposed disappearance at singularities, then singularities should be closely investigated. The least radical approach is usually considered the avoidance of singularity, by modifying gravity ({\ie} the relation between the stress-energy tensor and the spacetime curvature as expressed by the Einstein equation) so that one or more of the three assumptions of the singularity theorems \cite{Pen65,HP70,HE95} no longer hold. In particular, it is hoped that this may be achieved by the quantum effects in a theory of quantum gravity. However, it would be even less radical if the problem could be solved without modifying general relativity, and such an approach is the subject of this section.

But singularities are accompanied by divergences in the very quantities involved in the Einstein equation, in particular the curvature and the stress-energy tensor. So even if it is possible to reformulate the Einstein equation in terms of variables that do not diverge, remaining instead finite at the singularity, the question remains whether the physical fields diverge or break down. In other words, what are in fact the true, fundamental physical fields, the diverging variables, or those that remain finite? This question will be addressed soon.

An earlier mention of the possibility of changing the variables in the Einstein equation was made by Ashtekar, for example in \cite{ASH91} and references therein, where it is also proposed that the new variables could remain finite at singularities even in the classical theory. However, it turned out that one of his two new variables diverges at singularities (see {\eg} \cite{Yon97}). Eventually this formulation led to loop quantum gravity, where the avoidance is instead achieved on some toy bounce models (see {\eg} \cite{Ashtekar2011BlackHoleEvaporation2D} and \cite{Rovelli2014PlanckStars}).
But the problem whether standard general relativity can admit a formulation free of infinities at singularities remained open for a while.

\subsection{Singular general relativity}
\label{s:sgr}

In \cite{Sto11a,Sto11b}, the author introduced a mathematical formulation of semi-Riemannian geometry which allows a description of a class of singularities free of infinities. The fields which allowed this are invariant, and in the regions without singularities they are equivalent to the standard formulation. To understand what is the problem and how is solved, recall that in geometry the metric tensor is assumed to be smooth and regular, that is, without infinite components, and non-degenerate, which means that its determinant is nonvanishing. If the metric tensor has infinite components or if it is degenerate, the metric is called singular. If the determinant is vanishing, one cannot define the Levi-Civita connection, because the definition relies on the Christoffel symbols of the second kind,
\begin{equation}
\label{eq:christoffel}
\Gamma^i_{jk} := \frac 1 2 g^{is}\(g_{sj,k} + g_{sk,j} - g_{jk,s}\),
\end{equation}
which involve the contraction with $g^{is}$, which is the inverse of the metric tensor $g_{ij}$, hence it assumes it to be nondegenerate. This makes impossible to define the covariant derivative and the Riemann curvature (hence the Ricci and scalar curvatures as well) at the points where the metric is degenerate. These quantities blow up as approaching the singularities. Therefore, Einstein's equation as well breaks down at singularities.

However, it turns out that on the space obtained by factoring out the subspace of isotropic vectors, an inverse can be defined in a canonical and invariant way, and that there is a simple condition which leads to a finite Riemann tensor which is defined smoothly over the entire space, including at singularities. This allows the contraction of a certain class of tensors, and the definition of all quantities of interest to describe the singularities without running into infinities, and is equivalent to the usual, non-degenerate semi-Riemannian geometry, outside the singularities \cite{Sto11a}. Moreover, it works well for warped products \cite{Sto11b}, allowing the application to big bang models \cite{Sto11h,Sto12a}. This approach also works for black hole singularities \cite{Sto11e,Sto11f,Sto11g}, allowing the spacetime to be globally hyperbolic even in the presence of singularities \cite{Sto12e}. More details can be found in \cite{Sto13a,Sto14c} and the references therein. Here I will first describe some of the already published results, and continue with new and more general arguments.

An essential difficulty related to singularities is given by the fact that, despite the Riemann tensor being smooth and finite at such singularities, the Ricci tensor $R_{ij}:=R^s_{isj}$ usually continues to blow up. The Ricci tensor, as well as its trace, the scalar curvature $R=R^s_s$, are necessary to define the Einstein tensor, $G_{ij}=R_{ij}-\frac 1 2 R g_{ij}$. Now here is the part where the physical interpretation becomes essential. In the Einstein equation, the Einstein tensor is equated to the stres-energy tensor. So they both seem to blow up, and indeed they do. Physically, the stress-energy tensor represents the density of energy and momentum at a point. However, what is physically measurable is never such a density at a point, but its integral over a volume. The energy or momentum in a finite measure volume is obtained by integrating with respect to the volume element. And the quantity to be integrated, for example the energy density $T_{00}\de_{vol}$, where $T_{00}=T(u,u)$ for a timelike vector $u$ and $\de_{vol}:=\sqrt{-\det g}\de x^0\wedge\de x^1\wedge\de x^2\wedge\de x^3$, is finite, even if $T_{00}\to\infty$, since $\de_{vol}\to0$ in the proper way. The mathematical theory of integration on manifolds makes it clear that what we integrate are differential forms, like $T_{00}\de_{vol}$, and not scalar functions like $T_{00}$. So I suggest that we should do in physics the same as in geometry, because it makes more sense to consider the physical quantities to be the differential forms, rater than the scalar components of the fields \cite{Sto11h}. This is also endorsed by two other mathematical reasons. On the one hand, when we define the stress-energy $T_{ij}$, we do it by functional derivative of the Lagrangian with respect to the metric tensor, and the result contains the volume element. Which we then divide out to get $T_{ij}$. Should we keep it, we would get instead $T_{ij}\de_{vol}$. Also, when we derive the Einstein equation from the Lagrangian density $R$, we in fact vary the integral of the differential form $R\de_{vol}$, and not of the scalar $R$. And the resulting Einstein equation has again a factor $\de_{vol}$, which we leave out of the equation on the grounds that it is never vanishing. Well, at singularities it vanishes, so we should keep it, because otherwise we divide by $0$ and we get infinities. The resulting densitized form of the Einstein equation,
\begin{equation}
\label{eq:densitized-einstein}
G_{ij}\de_{vol} + \Lambda g_{ij}\de_{vol} = \frac{8\pi G}{c^4}T_{ij}\de_{vol}, 
\end{equation}
is equivalent to Einstein's outside singularities, but as already explained, I submit that it better represents the physical quantities, and not only because these quantities remain finite at singularities. I call this \emph{densitized Einstein equation}, but they are in fact tensorial as well, the fields involved are tensors, being the tensor products between other tensors and the volume form, which itself is a completely antisymmetric tensor. Note that Ashtekar's variables are also densities, and they are more different from the usual tensor fields involved in the semi-Riemannian geometry and Einstein's equation, yet they were proposed to be the real variables both for quantization and for eliminating the infinities in the singularities \cite{ASH91}. But the formulation I proposed remains finite even at singularities, and it is closer as interpretation to the original fields.

Another difficulty this approach had to solve was that it applies to a class of degenerate metrics, but the black hole are nastier, since the metric has components which blow up at the singularities. 
For example, the metric tensor of the {\schw} black hole solution, expressed in the {\schw} coordinates, is:
\begin{equation}
\label{eq_schw_schw}
\de s^2 = -\(1-\dsfrac{2m}{r}\)\de t^2 + \(1-\dsfrac{2m}{r}\)^{-1}\de r^2 + r^2\de\sigma^2,
\end{equation}
where $m$ is the mass of the body, the units were chosen so that $c=1$ and $G=1$, and
\begin{equation}
\label{eq_sphere}
\de\sigma^2 = \de\theta^2 + \sin^2\theta \de \phi^2
\end{equation}
is the metric of the unit sphere $S^2$.

For the horizon $r=2m$, the singularity of the metric can be removed by a singular coordinate transformation, see for example \cite{eddington1924comparison,finkelstein1958past}. Nothing of this sort could be done for the $r=0$ singularity, since no coordinate transformation can make the Kretschmann scalar $R^{ijkl}R_{ijkl}$ finite. However, it turns out that it is possible to make the metric at the singularity $r=0$ into a degenerate and analytic metric by coordinate transformations. In \cite{Sto11e} it was shown that this is possible, and an infinite number of solutions was found which lead to an analytic metric degenerate at $r=0$. Among these solutions, there is a unique one which satisfies the condition of semiregularity from \cite{Sto11a}, which ensure the smoothness and analyticity of the solution for the interior of the black hole. This transformation is
\begin{equation}
\label{eq_coordinate_semireg}
\begin{array}{l}
\left\{
\begin{array}{ll}
r &= \tau^2 \\
t &= \xi\tau^4 \\
\end{array}
\right.
\\
\end{array}
\end{equation}
and the resulting metric describing the interior of the {\schw} black hole is
\begin{equation}
\label{eq_schw_semireg}
\de s^2 = -\dsfrac{4\tau^4}{2m-\tau^2}\de \tau^2 + (2m-\tau^2)\tau^4\(4\xi\de\tau + \tau\de\xi\)^2 + \tau^4\de\sigma^2.
\end{equation}

This is not to say that physics depends on the coordinates. It is similar to the case of switching from polar to Cartesian coordinates in plane, or like the Eddington-Finkelstein coordinates. In all these cases, the transformation is singular at the singularity, so it is not a diffeomorphism. The atlas, the differential structure, is changed, and in the new atlas, with its new differential structure, the diffeomorphisms preserve, of course, the semiregularity of the metric. And just like in the case of the polar or spherical coordinates and the Eddington-Finkelstein coordinates, it is assumed that the atlas in which the singularity is regularized is the real one, and the problems were an artifact of the {\schw} coordinates, which themselves were in fact singular.

Similar transformations were found for the other types of black holes ({\rn}, Kerr, and {\kn}), and for the electrically charged ones the electromagnetic field also no longer blows up \cite{Sto11f,Sto11g}.

\subsection{Beyond the singularity}
\label{s:sgr-time-evol}

Returning to the {\schw} black hole in the new coordinates \eqref{eq_schw_semireg}, the solution extends analytically through the singularity. If we plug this solution in the Oppenheimer-Snyder black hole solution, we get an analytic extension depicting a black hole which forms and then evaporates, whose Penrose-Carter diagram is represented in fig. \ref{bh-info-analytic.png}.

\image{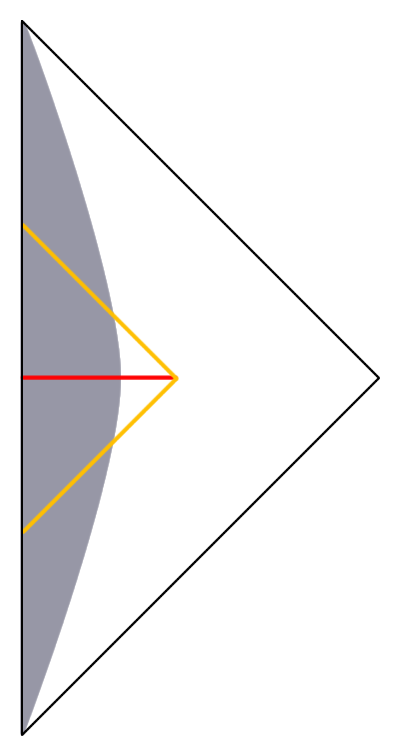}{0.3}{An analytic extension of the black hole solution beyond the singularity.}

The resulting spacetime does not have Cauchy horizons, being hyperbolic, which allows the partial differential equations describing the fields on spacetime to be well posed and continued through the singularity. Of course, there is still the problem that the differential operators in the field equations of the matter and gauge fields going through the singularity should be replaced with the new ones. Such formulations are introduced in \cite{Sto14b}, and sufficient conditions to be satisfied by the fields at the singularities so that their evolution equations work were given, in the case of Maxwell and Yang-Mills equations.

It is an open problem whether the backreaction will make the spacetime to curve automatically so that these conditions to be satisfied for all possible initial conditions of the field. This should be researched in the future, including for quantum fields. It is to be expected that the problem is difficult, and what is given here is not the general solution, but rather a toy model. Anyway, no one should expect very soon an exact treatment of real case situations, so the whole discussion here is in principle, to establish whether this conservative approach is plausible enough.

However, I would like to propose here a different, more general argument, which avoids the difficulties given by the necessity that the field equations should satisfy at the singularities special conditions like the sufficient conditions found in \cite{Sto14b}, and also the open problem of which are the conditions to be satisfied by the fermionic fields at singularities.

First consider Fermat's principle in optics. A ray of light in geometric optics is straight, but if it passes from one medium to another having a different refraction index, the ray changes its direction and appears to be broken. It is still continuous, but the velocity vector is discontinuous, and it appears that the acceleration blows up at the surface separating the two media. But Fermat's principle still allows us to know exactly what happens with the light ray in geometric optics.

On a similar vein, I think that in the absence of a proof that the fields satisfy the exact conditions \cite{Sto14b} when crossing a singularity, we can argue that the singularities are not a threat to the information contained in the field by using the least action principle instead.

The least action principle involves the integration of the Lagrangian densities of the fields. While the conditions the fields have to satisfy at the singularity in order to behave well are quite restrictive, the Lagrangian formulation is much more general. The reason is that integration can be done over fields with singularities, also on distributions, and the result can still be finite.

Consider first classical, point-like particles falling in the black hole, crossing the singularity, and exiting through the white whole which appears after the singularity disappears. The history of such a test particle is a geodesic, and to understand the behavior of geodesics, we need to understand first the causal structure. In fig. \ref{bh-null-geodesics.png}, the causal structure of \textbf{A.} a {\schw} black hole, and \textbf{B.} a {\rn} black hole, are represented in the coordinates which smoothen the singularity (see \cite{Sto15b}).

\image{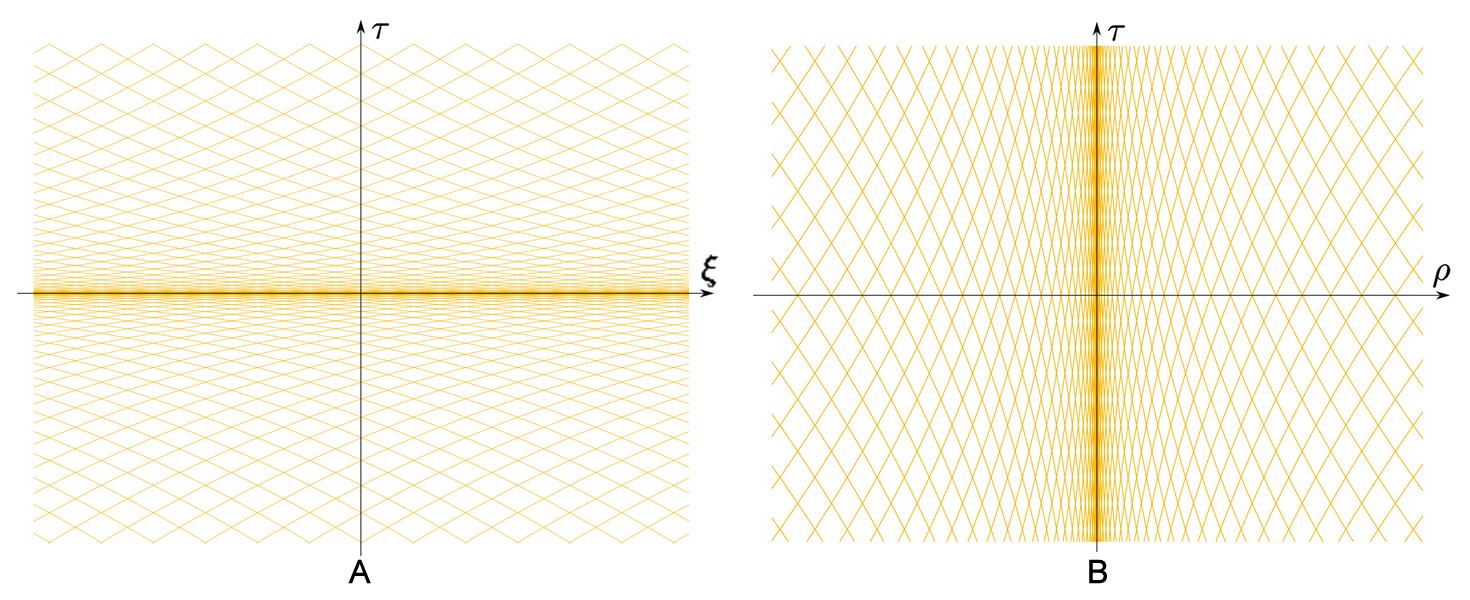}{0.99}{\textbf{A.} The causal structure of the {\schw} black hole in coordinates $(\tau,\xi)$ from equation \eqref{eq_coordinate_semireg}. \textbf{B.} The causal structure of the {\rn} black hole, in coordinates $(\tau,\rho)$ playing a similar role, see \cite{Sto11f}.}

If the test particle is massless, its path is a null geodesic. In \cite{Sto15b} I showed that for the standard black holes, the causal structure at singularities is not destroyed. The lightcones will be squashed, but they will remain lightcones. Therefore, the history of a massless particle like a photon is, if we apply the least action principle, just a null geodesic crossing the singularity and getting out.

If the test particle is massive, its history is a timelike geodesic. In this case a difficulty arises, because in the new coordinates the lightcones are squashed. This allows for distinct geodesics to intersect the singularity at the same point and to have the same spacetime tangent direction. In the {\schw} case this does not happen for timelike geodesics, but in the {\rn} case \cite{Sto11f} all of the timelike geodesics crossing the singularity at the same point become tangent. Apparently, this seems to imply that a geodesic crossing a timelike singularity can get out of it in any possible direction, in a completely undetermined way. To fix this, one may want to also consider the second derivative, or to use the local cylindrical symmetry around the timelike singularity.

But the least action principle allows this to be solved regardless of the specific local solution of the problem at the singularity. The timelike geodesics are tangent only at the singularity, which is a zero-measure subset of spacetime. So we can apply the least action principle to obtain the history of a massive particle, and obtain a unique solution. The least action principle can be applied for classical test particles because a particle falling in the black hole reaches the singularity in finite proper time, and similarly a finite proper time is needed for it to get out. Moreover, the path integral quantization will consider anyway all possible paths, so even if there would be an indeterminacy at the classical level, it will be removed by integrating them all.

For classical fields the same holds as for point-like classical particles, only the paths are much more difficult to visualize. The least action principle is applied in the configuration space even for point-like particles, and the the same holds for fields, the only difference being the dimension of the configuration space and the Lagrangian. The points from the singularity form again a zero-measure subset compared to the full configuration space, so finding the least action path is similar to the case of point-particles. 
The Lagrangian density is finite at least at the points of the configuration space outside the singularities, which means almost everywhere. But the volume element vanishes at singularities, which improves the situation. So its integral can very well be finite, even if the Lagrangian density would be divergent at the singularities. It may be the case that the fields have singular Lagrangian density at the singularity, and that when we integrate them it is not excluded that even the integral may diverge, but in this case the least action principle will force us anyway to choose the paths which have a finite action density at the singularities, and such paths exist, for example those satisfying the conditions found in \cite{Sto14b}. 

So far we have seen that the principle of least action allows to determine the history of classical, point-like particles or fields, from the initial and final conditions, even if they cross the singularity. This is done so far on fixed background, so no backreaction via Einstein's equation is considered, only particles or fields. But the Lagrangian approach extends easily to include the backreaction, we simply add the Hilbert-Einstein Lagrangian to that of the fields or point-particles. So now we not only vary the path of point-particles or fields in the configuration space, but also the geometry of spacetime, in order to find the least action history. This additional variation gives even more freedom to choose the least action path, so even if on fixed background the initial condition of a particular field will not evolve to become, at the singularity, a field satisfying the conditions from \cite{Sto14b}, because the spacetime geometry is varied as well to include backreaction, the spacetime adjusts itself to minimize the action, and it is not too wild to conjecture that it adjust itself to satisfy such conditions.

Now let us consider quantum fields. 
When moving to quantum fields on curved background, since the proper time of all classical test particles is finite, we can apply the path integral formulation of quantum field theory \cite{Dirac1933PathIntegral,FeynmanHibbs2012QMAndPathIntegrals}.
Since the proper time is finite along each path $\varphi$ joining two points, including for the paths crossing a singularity, and since the action $S(\varphi,t)$, is well defined for almost all times $t$, then $e^{\frac{i}{\hbar}S(\varphi,t)}$, is also well defined. So at least on fixed curved background, even with singularities, it seems to be little difference from special relativistic quantum field theory via path integrals.

Of course the background geometry should also depend on the quantum fields. Can we account for this, in the absence of a theory of quantum gravity? We now that at least the framework of path integrals works on curved classical spacetime (see {\eg} \cite{Kleinert2009PathIntegrals}), where the Einstein equation becomes \eqref{eq:einstein-qft}. To also include quantized gravity is more difficult, because of its nonrenormalizability by perturbative methods. Add to this the fact that at least for the Standard Model we know that in flat background renormalization helps, and even on curved background without singularities. But what about singularities, isn't it possible that they make renormalization impossible? In fact, quite the contrary may be true: in \cite{Sto12d} it is shown that singularities improve the behavior of the quantum fields, including for gravity, at UV scales. These results are applied to already existing results obtained by various researchers, who use various types of dimensional reduction to improve this behavior for quantum fields, including gravity. In fact, some of these approaches improve the renormalizability of quantum fields so well that even the Landau poles disappear even for nonrenoramlizable theories \cite{FS2011KG,shirkov2012dreamland}. But the various types of dimensional reduction are, in these approaches, postulated somehow \emph{ad hoc}, for no other reason than to improve perturbative renormalizability. By contrary, if the perturbative expansion is made in terms of point-particles, these behave like black holes with singularities, and some of the already postulated types of dimensional reduction emerge automatically, with no additional assumption, from the properties of singularities \cite{Sto12d}. Thus, the very properties of the singularities lead automatically to improved behavior at the UV scale, even for theories thought to be perturbatively nonrenormalizable.

The proposal I described in this section is still at the beginning, compared to the difficulty of the remaining open problems to be addressed. First, there is obviously no experimental confirmation, and it is hard to imagine that the close future can provide one. The plausibility rests mainly upon making as few new assumptions as possible, in addition to those coming from general relativity and quantum theory, theories well established and confirmed, but not in the regimes where both become relevant. For some simple examples there are mathematical results, but a truly general proof, with fully developed mathematical steps and no gaps, does not exist yet. And considering the difficulty of the problem, it is hard to believe that it is easy to have very soon a completely satisfying proof, in this or other approaches. Nevertheless, I think that promising avenues of research are opened by this proposal.

\end{document}